\documentclass{aa}  
\usepackage{graphicx}
\usepackage{epsfig,color}
\usepackage{txfonts}
\usepackage{natbib}
\bibpunct{(}{)}{;}{a}{}{,}
\begin{document}
\def\etal{{\rm et al.\thinspace}}
\def\eg{{\rm e.g.}}
\def\etc{{\rm etc.\ }}
\def\ie{{\rm i.e.\ }}
\def\cf{{\rm cf.\ }}
\def\perse{{\it per se}}
\def\spose#1{\hbox to 0pt{#1\hss}}
\def\ltsimm{\mathrel{\spose{\lower 3pt\hbox{$\sim$}}
	\raise 2.0pt\hbox{$<$}}}
\def\gtsimm{\mathrel{\spose{\lower 3pt\hbox{$\sim$}}
	\raise 2.0pt\hbox{$>$}}}
\def\Mdot{\hbox{${\dot M}$}}
\def\vinfty{\hbox{${v_{\infty}}$} \,}
\def\km{{\rm\thinspace km}}
\def\cm{{\rm\thinspace cm}}
\def\pc{{\rm\thinspace pc}}
\def\s{{\rm\thinspace s}}
\def\yr{{\rm\thinspace yr}}
\def\g{{\rm\thinspace g}}
\def\K{{\rm\thinspace K}}
\def\massh{{\rm\thinspace m_{H}}}
\def\kmps{\hbox{${\rm\km\s^{-1}\,}$}}
\def\erg{{\rm\thinspace erg}}
\def\ergps{\hbox{${\rm\erg\s^{-1}\,}$}}
\def\Msol{\hbox{${\rm\thinspace M_{\odot}}$}}
\def\Msolpyr{\hbox{${\rm\Msol\yr^{-1}\,}$}}
\def\ergpcm2ps{\hbox{${\rm\erg\cm^{-2}\s^{-1}\,}$}}
\def\pcm3{\hbox{${\rm\cm^{-3}\,}$}}
\def\gpcm3{\hbox{${\rm\g\cm^{-3}\,}$}}
\def\gpcm3ps{\hbox{${\rm\g\cm^{-3}\s^{-1}\,}$}}
\def\gps{\hbox{${\rm\g\s^{-1}\,}$}}

\title{The tails in the Helix Nebula NGC 7293}

\author{J.E. Dyson\inst{1}, J.M. Pittard\inst{1}, J. Meaburn\inst{2} \and S.A.E.G. Falle\inst{3}}

\offprints{J.E. Dyson}

\institute{
School of Physics and Astronomy, University of Leeds, Leeds LS2 9JT\\
\email{jed@ast.leeds.ac.uk}
\and 
Jodrell Bank Observatory, University of Manchester, Macclesfield SK11 9DL
\and Department of Applied Mathematics, University of Leeds, Leeds LS2 9JT}

\date{Received ....; accepted ....}

  \abstract
   {}
   {We have examined a stream-source model for the production of the
   cometary tails observed in the Helix Nebula NGC~7293 in which a
   transonic or moderately supersonic stream of ionized gas overruns a
   source of ionized gas. We have compared the velocity structures
   calculated with the available observational data. We have also
   investigated the suggestion that faint striations visible in the
   nebular gas are the decaying tails of now destroyed cometary
   globules.}
   {We have selected relevant results from extensive hydrodynamic
   calculations made with the COBRA code.}
   {The velocities calculated are in good agreement with the
   observational data on tail velocities and are consistent with
   observations of the nebular structure. The results also are
   indicative of a stellar atmosphere origin for the cometary
   globules. Tail remnants persist for timescales long enough for
   their identification with striations to be plausible.}
   {}


\maketitle

\section{Introduction}
Many PNe contain extended cometary tail-like structures. These emanate
from dense, neutral \citep{Meaburn:1992,Huggins:1992} globules with
radiatively ionized surfaces and point away, comet-like, from the
central star. The most studied are those in the highly evolved PNe
NGC~7293 (the Helix Nebula). A brief review of the two broad classes
of models proposed to account for them is given by Dyson
(2003). ``Shadow'' models \citep[e.g.,][]{VanBlerkom:1972,Canto:1998,
Lopez-Martin:2001} use an opaque cloud to shield gas from the direct
radiation field of the PNe nuclear star. The shadow region is
energized by a diffuse radiation field of different quality from the
direct stellar field. ``Stream-source'' models (\citet{Dyson:1993} - henceforth
Paper~I, \citet{Falle:2002} - henceforth Paper~II) are
based on mass injection from embedded clumps in global flows. There
are three relevant length scales \citep{Hartquist:1993}.  The smallest
is the mass injection scale and the largest scale is that for
assimilation of injected gas into the global flow. The intermediate
scale where injected material largely retains its own identity,
produces a tail feature that is dynamically affected by momentum
transfer from the global flow.

\citet{Lefloch:1994} argued for tail formation during the
photoionization of dense globules, a suggestion also made by
\citet{Rijkhorst:2006} who did not, however, present any models.  Yet
as shown by \citet{Pavlakis:2001}, the diffuse radiation field which
was neglected by \citet{Lefloch:1994} can suppress tail formation if
it has an intensity of a few percent of the direct stellar flux.
Given that there are several thousand tails distributed throughout the
nebula with concommitant variations in direct flux, diffuse flux and
inevitably globule properties, it is hard to see how the tail system
could be produced by this mechanism.  Stream-source models also have
the advantage that they are inherently dynamic in distinction to the
majority of shadow models where the dynamics of the gas is neglected,
though the model of \citet{Lopez-Martin:2001} includes a flow of
ionized gas perpendicular to a tail feature as the gas exits from a
D-critical ionization front at the ionized sound speed. However, the
only detailed observational study of the velocity structure of the
Helix tails by \citet{Meaburn:1998} shows no evidence for significant
ionized gas velocities perpendicular to the tails.

The ratio of tail length to head diameter of the Helix cometary
features ranges from $\approx 1$ to $\approx 100$ in the spectacular
knot~38 \citep{Meaburn:1998} which has a deprojected actual tail length of
$\approx 2.5 \times 10^{17}\;{\rm cm}$. Tails are aligned to within a
few degrees along radial vectors originating at the central star. Our
discussions are directed specifically at the Helix tails and clumps
and their generalisation to other PNe may be inappropriate.

In this paper we re-examine the stream-source models proposed in
Paper~I using extensive calculations of stream-source interactions by
Dyson et al. (in preparation - henceforth Paper IV). In Paper~I, a hot
subsonic stream of bottled-up shocked stellar wind gas was assumed to
interact with a source of cooler gas to produce tails. Serious
objections to this tail generation mechanism were raised by
\citet{ODell:2000} and \citet{Meaburn:2005}. The main problem is that
the central cavity in the Helix is occupied by a hot ($T_{\rm e}
\gtsimm 2 \times 10^{4}\;{\rm K}$) He$^{+}$ emitting gas and it is
hard to see where the hot shocked stellar wind gas resides. In fact
much earlier, \citet{Meaburn:1982} had raised a similar objection
because of the presence of an inner [OIII] emitting shell that would
shield the cometary globules from the direct impact of any fast
(subsonic or supersonic) wind.

Therefore we examine the suggestion that the cometary tail heads are
overrun by a transonic or supersonic stream of photoionized gas
\citep{Meaburn:1992,Meaburn:1998,ODell:2000,Meaburn:2005}. 
We compare results with the knot dynamics observed by
\citet{Meaburn:1998} and show these are reasonably well
reproduced. We briefly discuss the significance of clump motions for
these models.

\citet{Meaburn:2005} noted the existence of ``headless'' radial spokes
close to the central star of NGC~7293 and suggested that these spokes
were tails left behind after their cometary knots were photoevaporated 
away. \citet{Henry:1999} and \citet{ODell:2004} also noted
similar striations in the nebular gas. We examine the effects of
switching off the mass injection and show that tails persist for
typical timescales of thousands of years before they are advected away
by the stream flow, lending plausibility to the suggested origin
\citep{Meaburn:2005}.

\section{Stream-source tail production}
Paper~I argued that long thin tails are produced only when subsonic
streams interact with subsonic mass injection (all Mach numbers are
defined in the reference frame of the clump). Paper~II dealt with the
interactions of hypersonic and transonic streams with injected matter.
\citet[][henceforth Paper~III]{Pittard:2005} investigated the
interaction of hypersonic flows with multiple clumps and the
interaction of transonic flows with two adjacent clumps. We will
consider only single clump interactions although both photographs of
tails and the recent estimate of 23000 cometary knots
\citep{Meixner:2005} suggest multiple clump interactions might occur
in the Helix.

In Papers~I~to~III it was assumed that there was a large temperature
contrast between a hot global flow and injected gas with the stream
gas behaving adiabatically and the injected gas behaving
isothermally. Tail dynamics and morphology are determined by the
thermal behaviour and temperature contrast of the stream and source
gas as well as by the Mach number of the incident stream. Paper~IV
investigates a broader range of incident Mach numbers and temperature
contrasts using spherical clump geometry (Paper~III shows that the
differences between spherical and cylindrical geometry are small). 
Results are also given for the case where both
stream and source gases behave isothermally. This is appropriate if
the stream gas as well as the source gas is photoionized and we
utilise these results here. Computational details are given in
Papers~III and~IV.

An important result from these new calculations is that identifiable
long thin tails persist to appreciably higher Mach number interactions
than suggested in Papers~I and~II. This is quantified in Paper~IV.

Mass loss from the clumps in the Helix is due to
photoionization. Clump images show the mass injection is strongest in
the direction towards the central star. In Paper~III it was shown that
anisotropic mass injection made very little difference to the flow
morphology in the case of a hypersonic wind interacting with source
material. Comparison of tail structures for subsonic and transonic
stream interactions where there is anisotropic injection is given in
Paper~IV. Results for a mass injection rate contrast of 20:1 (front to
rear) show small differences in the tail morphologies and tail
velocity structures between the isotropic and anisotropic cases once
the flow has gone a few injection radii. We use results for isotropic
mass injection here.

Although the models of Paper~IV are far more extensive than previously
given, they have still a major simplification.  They contain only two
gas phases (stream and injected gas) which may mix. The Helix tails
possibly represent a three phase system that includes ionised stream
gas and both neutral and ionised injected gas
\citep[e.g.,][]{ODell:1996,ODell:2000}. To treat this properly
requires the inclusion of radiation transfer.

\section{Tail production}
\citet{Meaburn:1998}, \citet{ODell:2000} and \citet{Meaburn:2005} have
shown that it is likely that clumps are overrun by photoionized gas
that may be ionized AGB wind possibly contaminated with evaporated
knot gas. We consider tail formation in an interaction where
photoionized gas overruns clumps that lose material by
photoionization. An isothermal-isothermal assumption for the thermal
behaviour of the stream and source gas is appropriate.

It is unlikely in general that the clumps are being overrun by the
very hot He$^{++}$ gas since the clumps have an average outward velocity of
$14\;\kmps$ \citep{Meaburn:1998} and the hot gas is dynamically inert
with an expansion velocity of less than $11\;\kmps$
\citep{Meaburn:2005}.  On the other hand, there might be clumps that
are overrun by this hot gas since the velocity dispersion of the knots
is around $6-8\;\kmps$ \citep{Meaburn:1998}. The temperature contrast
between the stream gas and injected gas may be
high. \citet{ODell:1998} suggests that the temperature of the
He$^{++}$ gas is in excess of $2 \times 10^{4}\;{\rm K}$ (because of
the lack of O$^{++}$ cooling). \citet{Henry:1999} have modelled the
ionization structure of the nebula and shown that the gas temperature
drops from about $4 \times 10^{4}\;{\rm K}$ near the star to
$10^{4}\;{\rm K}$ at a distance where helium is about 50\% in the form
of He$^{++}$.

\citet{Meaburn:1998} and \citet{Meaburn:2005} favour the overrunning
of clumps by expanding photoionized gas in the region where [OIII] is
emitted. This gas has a temperature of about $10^{4}\;{\rm K}$
\citep{Henry:1999}, and is overrunning the clumps with a velocity
of about $17\;\kmps$ relative to the clump global expansion velocity
of $14\;\kmps$.

The temperature of gas injected from the source is uncertain.
\citet{ODell:2005} note that the temperature of photoionized gas
flowing from a clump towards the star decreases outwards from the
density peak just behind the ionisation front on the clump
surface. Some injection will occur from the sides of clumps where the
diffuse radiation field softens the average photon energy leading to
gas temperatures that could be a factor of around 2 lower than the
temperature of gas excited by the direct radiation field
\citep{VanBlerkom:1972,Canto:1998}.

\begin{figure}
\centering
\psfig{figure=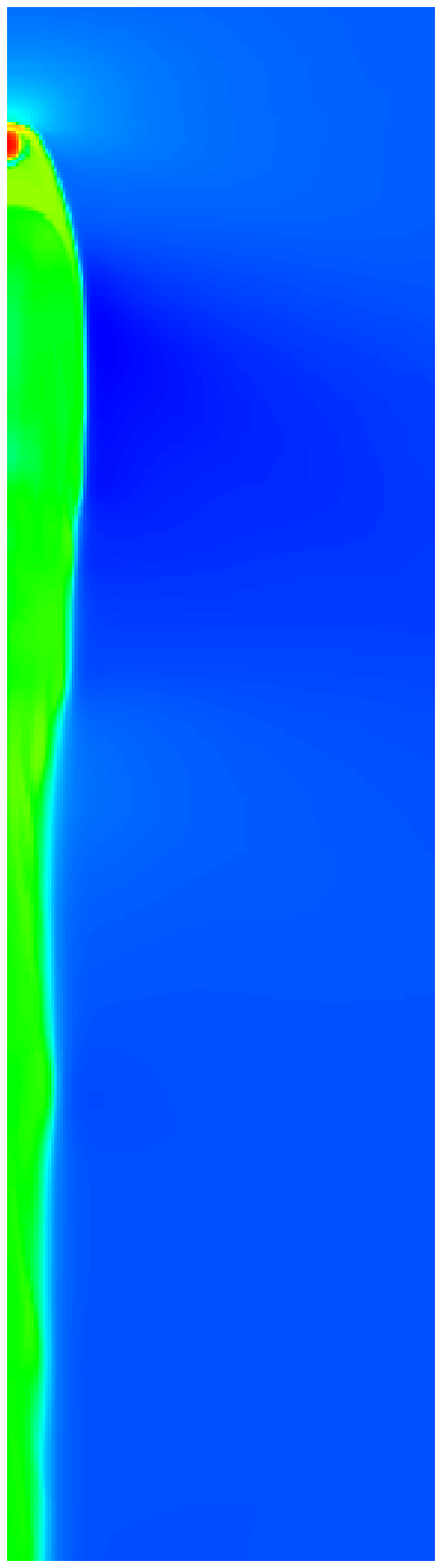,width=2.8cm}
\psfig{figure=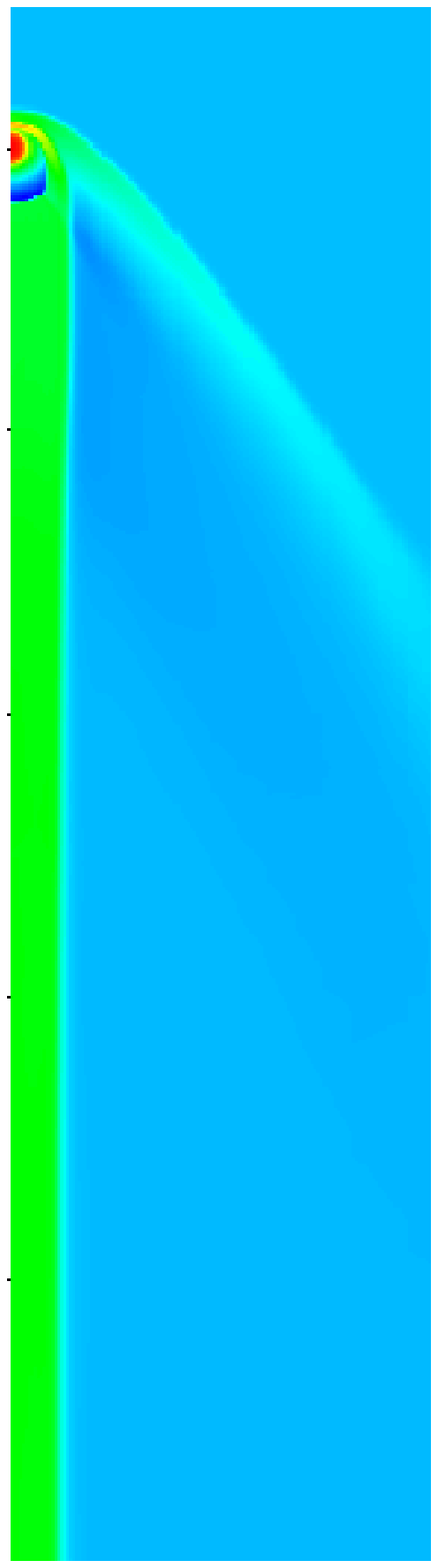,width=2.8cm}
\psfig{figure=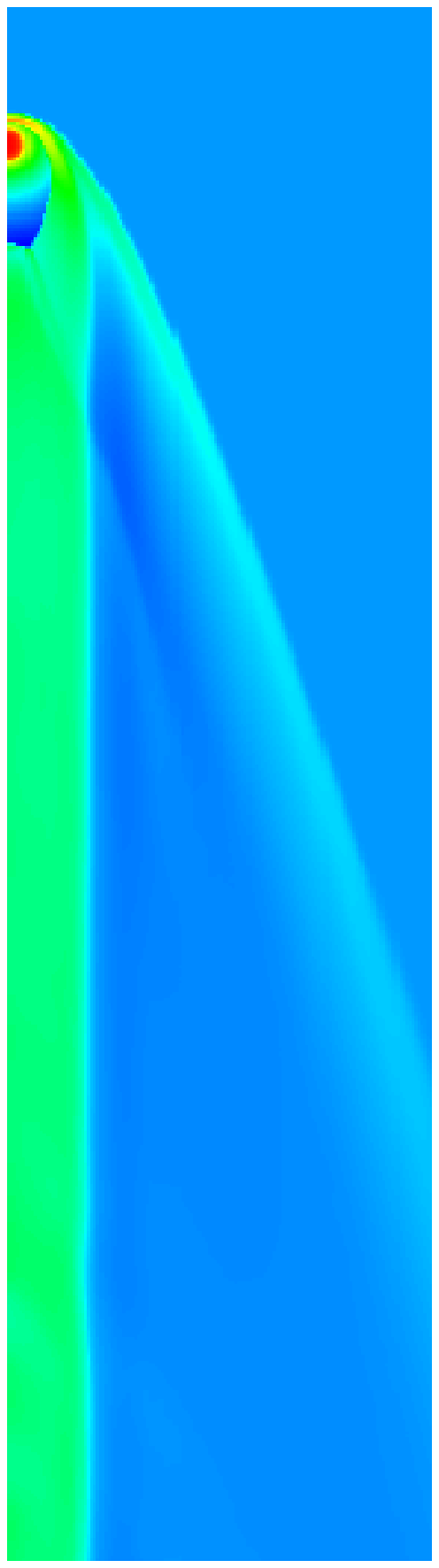,width=2.8cm}
\caption{Tail density structures for sound speed ratio 
$c_{\rm S}/c_{\rm i}=$2:1. The stream Mach number, $M=1,2,4$ for the left,
middle, and right panels respectively. Each panel shows a region
$0<r<30$, $-100<z<10$. The logarithm of mass density is shown, with a scale of
$-0.69$ to 0.39 (left), $-0.97$ to 0.99 (middle), and $-0.99$ to 1.60 (right).
Blue denotes the lowest density, while red denotes the highest. The stream
density is $0.25$. An injection region of unit radius is at the origin. 
See Paper II for further details about the computations.}
\label{fig:rho1}
\end{figure}

As a reasonable range for the temperatures we use calculations from
Paper~IV for ratios of isothermal sound speeds ($c_{\rm S}/c_{\rm i}$,
stream gas to injected gas) of 1:1 (equal temperatures) and 2:1 (a
temperature ratio of 4:1). In Fig.~\ref{fig:rho1} we give tail density
structures for a sound speed ratio of 2:1 for incident stream Mach
numbers $M=1,2,4$.

\begin{figure}
\centering
\psfig{figure=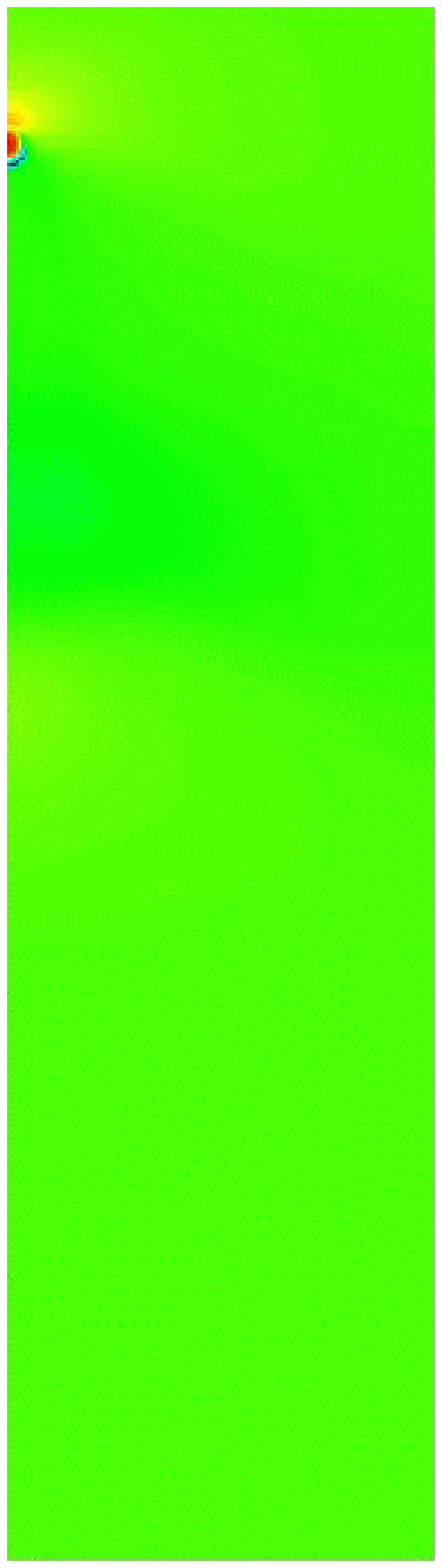,width=2.8cm}
\psfig{figure=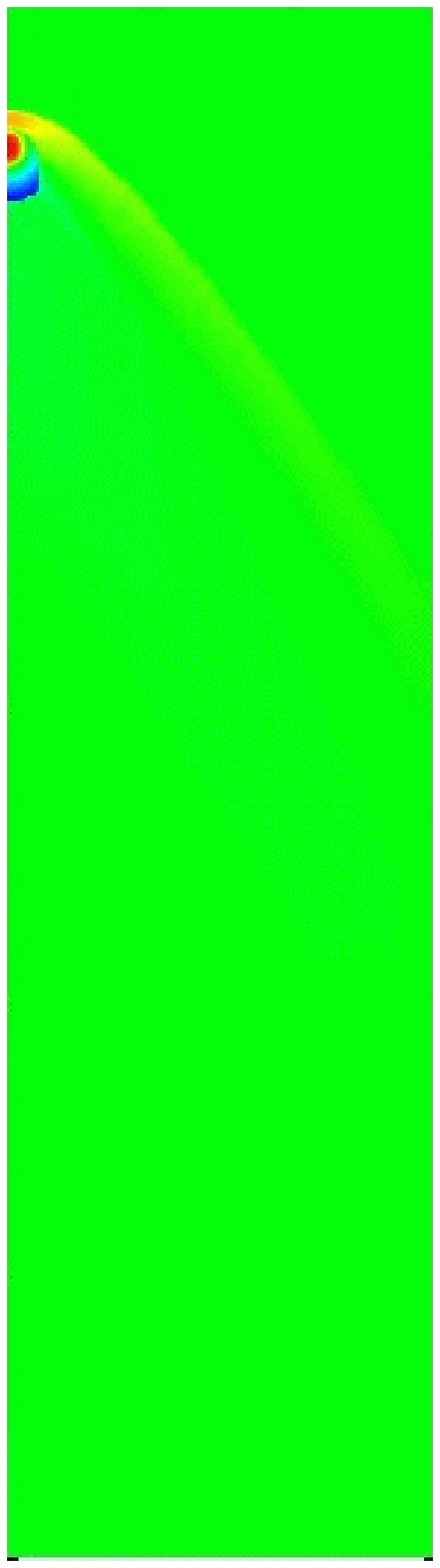,width=2.8cm}
\psfig{figure=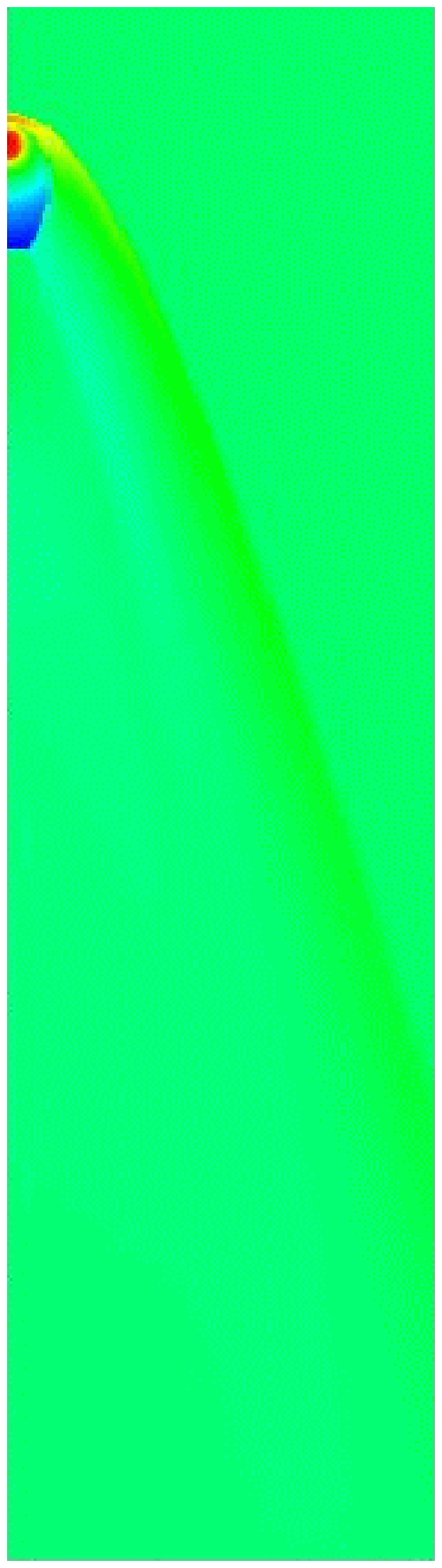,width=2.8cm}
\caption{As Fig.~\ref{fig:rho1} but for equal sound speeds with $M = 1,2,4$ 
(left, middle, and right, respectively). The density scale extends from $-0.52$
to 0.39 (left), $-0.97$ to 0.99 (middle), and $-1.02$ to 1.59 (right).
The stream density is 1.0.}
\label{fig:rho2a}
\end{figure}

\begin{figure}
\centering
\psfig{figure=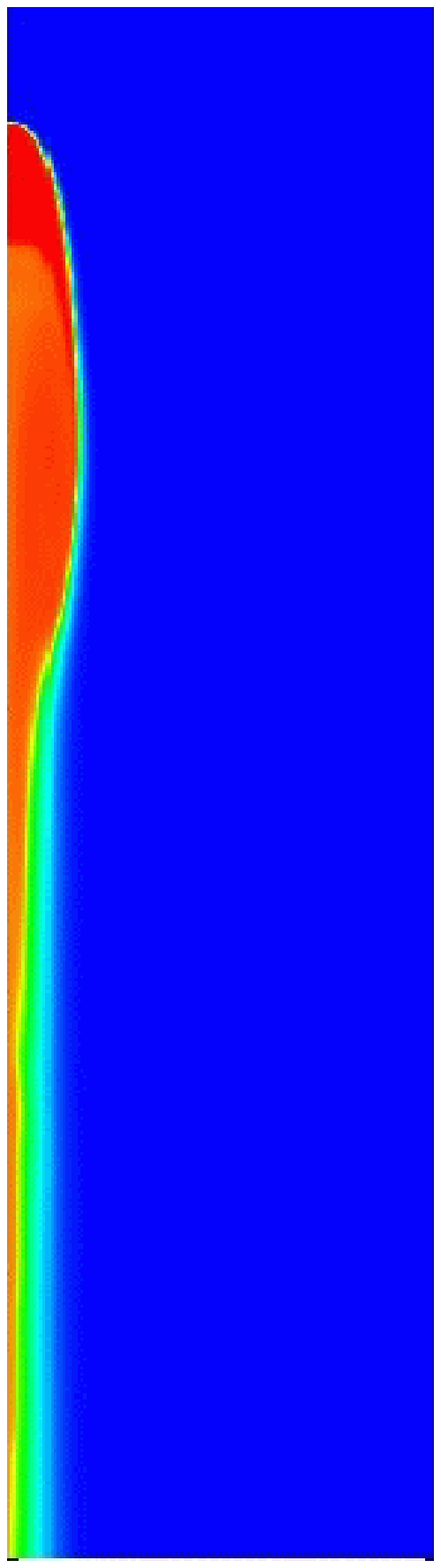,width=2.8cm}
\psfig{figure=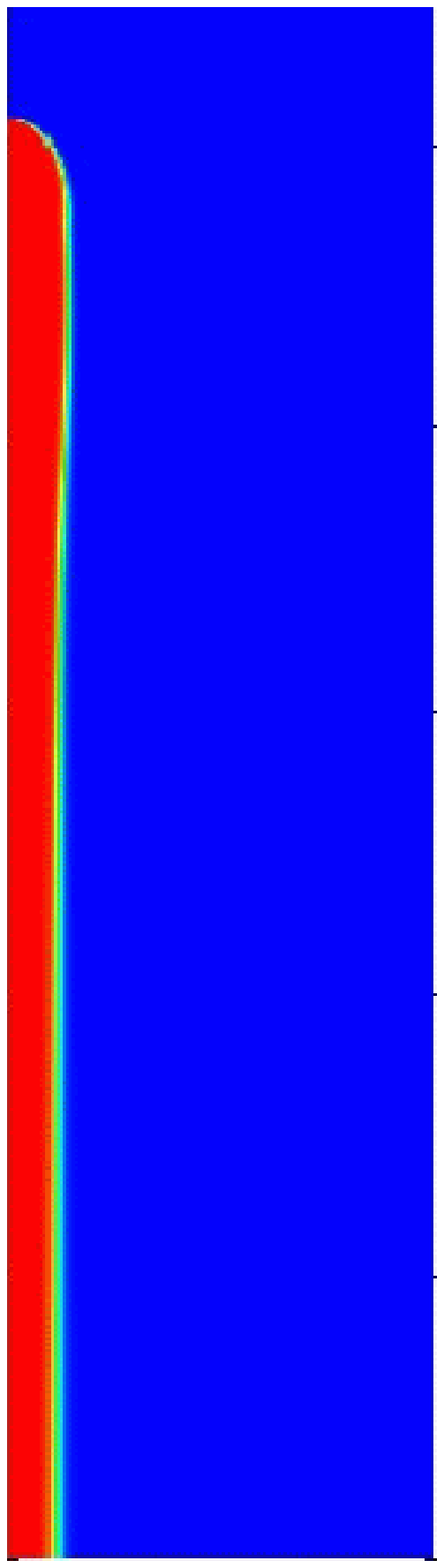,width=2.8cm}
\psfig{figure=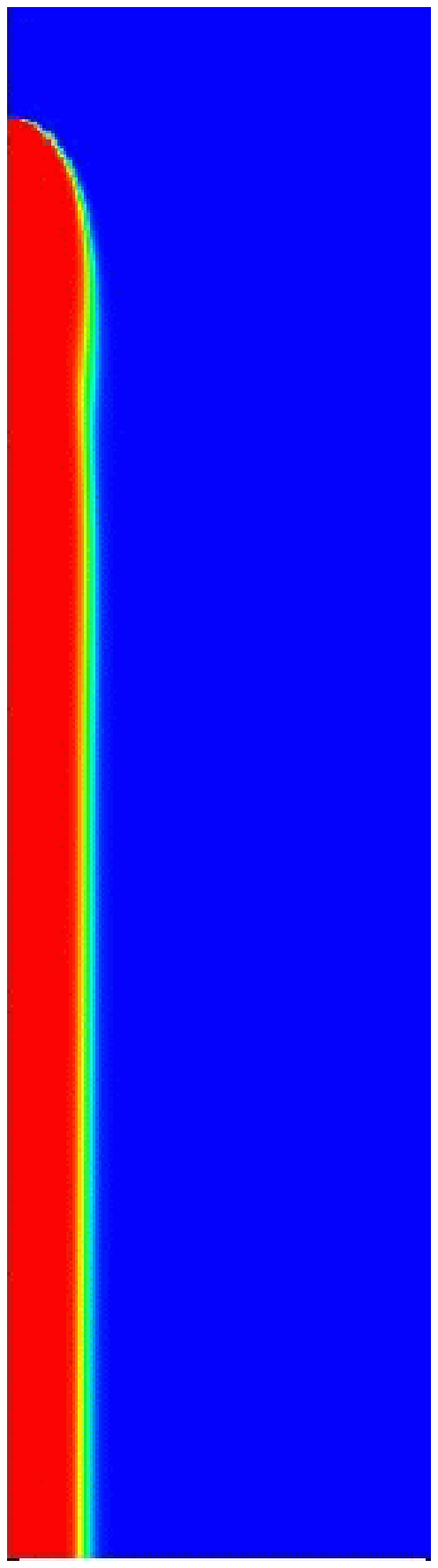,width=2.8cm}
\caption{As Fig.~\ref{fig:rho2a} but showing advected scalar structures.
Blue represents the stream gas, while red represents injected
material (i.e. material evaporated from the clump into the flow).}
\label{fig:rho2b}
\end{figure}

In Fig.~\ref{fig:rho2a} we give the same information for a sound speed
ratio 1:1. However a sound speed ratio of unity gives a tail density
effectively the same as that of the stream and the plots of
Fig.~\ref{fig:rho2a} therefore do not show systematic tail structure.
The results of Paper~IV demonstrate that the injected gas maintains
its identity and there is very little mixing between it and stream
gas. So to show this, we plot in Fig.~\ref{fig:rho2b} the advected
scalar that distinguishes between the stream and injected gases. A
recognisable tail of injected gas is embedded in stream gas. Some
difference in emission or absorption properties (e.g., a different
gas-to-dust ratio) is needed to make the tail visible as is clear from
the density plot but it is a dynamical possibility for tail
production.

\section{Tail velocity structures}
\label{sec:tailvz}
Important diagnostic information comes from the tail velocity
structure in knot~38 given by \citet{Meaburn:1998}. In view of our
model limitations, we use mass-weighted velocities for comparison with
the data.

\begin{figure}
\centering
\psfig{figure=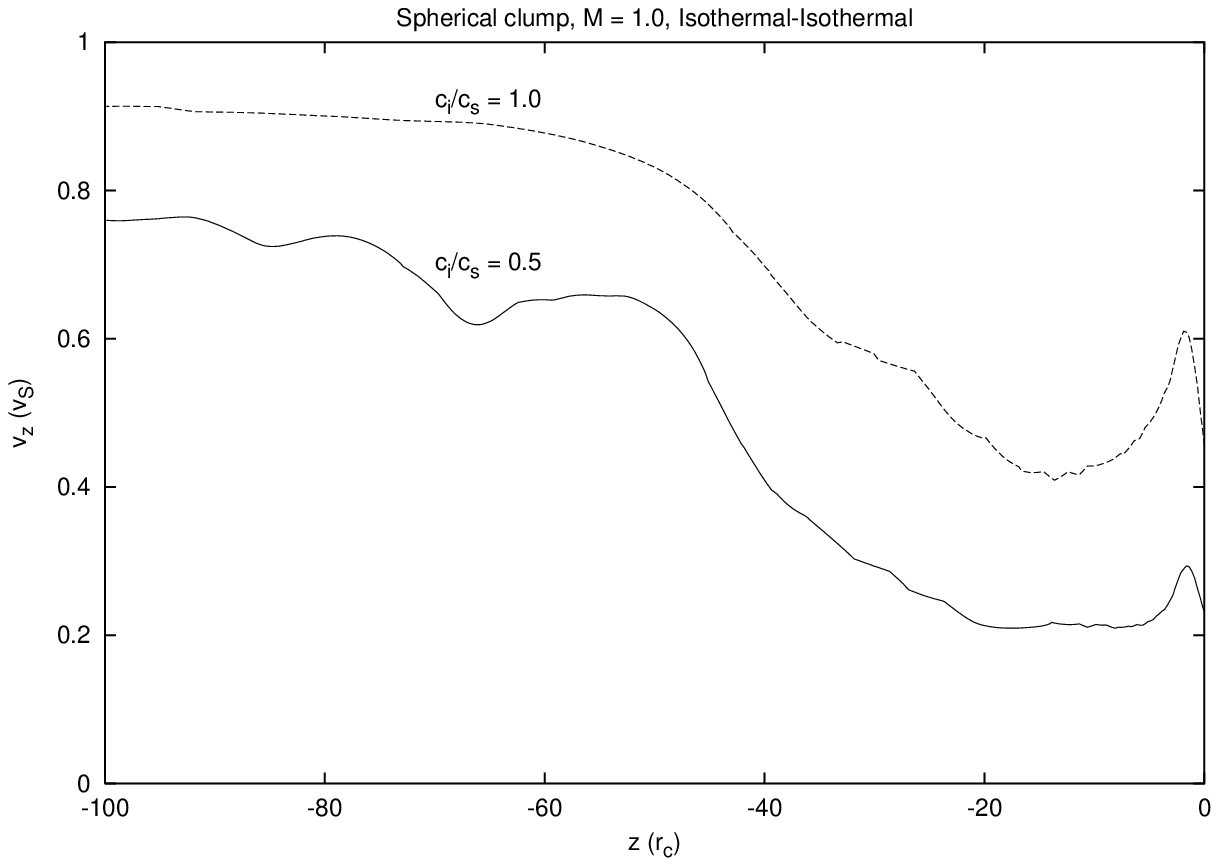,width=8.0cm}
\psfig{figure=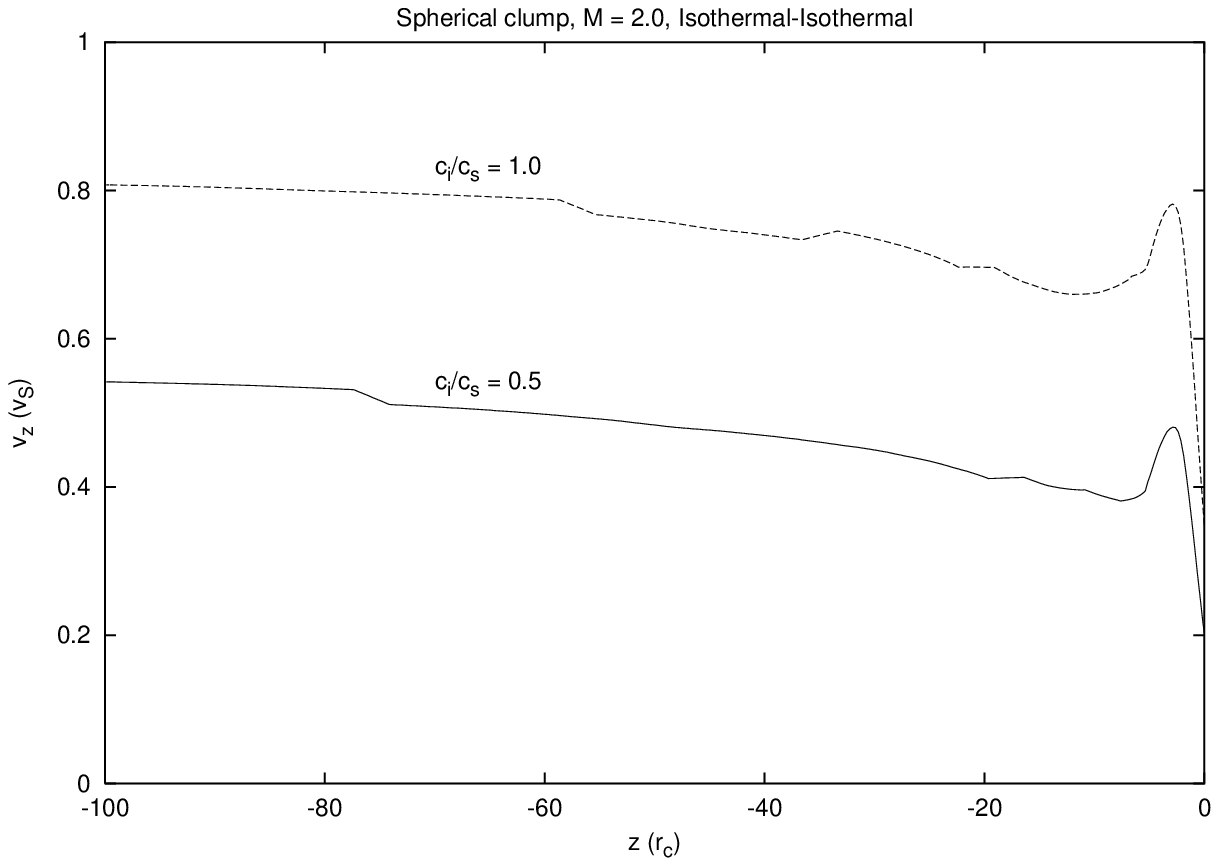,width=8.0cm}
\psfig{figure=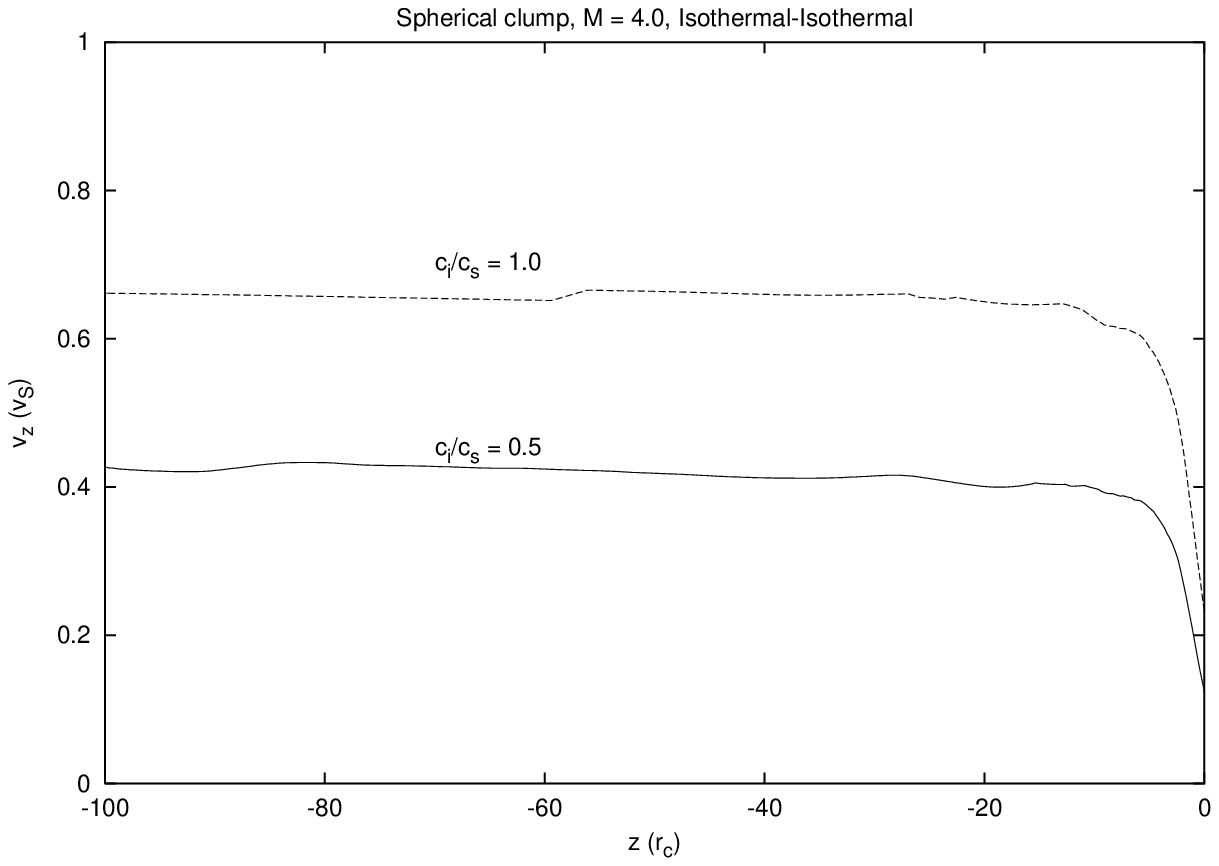,width=8.0cm}
\caption{$V_{\rm Z}$ vs $z$ for $M = 1,2,4$, and sound speed ratios 1 and 2 
(i.e. temperature ratios 1 and 4). The injection radius is $r_{\rm c}$.}
\label{fig:vz}
\end{figure}

In Fig.~\ref{fig:vz} we show mass-weighted velocity profiles along tails for
several incident Mach numbers and two stream-source gas temperature
ratios. We give results only for the transonic and supersonic
isothermal-isothermal flows that are relevant to the present paper.

For streams that are roughly in excess of transonic, the velocity
increases quickly over a few injection radii. The initial increase is
followed by a velocity decrease for Mach numbers less than about 3
then the velocity rises to a roughly coasting velocity. Once the Mach
number is around 3, the initial increase is just followed by a very
gentle rise or even roughly constant velocity over distances of tens
of injection radii. Appreciably greater distances are probably not
relevant to the Helix tails where the very longest tail (knot~38) has
a length about 100 times its head diameter and we display results
accordingly.

The final velocity reached depends on the temperature contrast between
the stream and source gas and is higher (in terms of the stream
velocity, $V_{\rm S}$) the lower the temperature contrast. This is as expected
since the temperature contrast determines crudely the stream-injected
gas density contrast and there is more momentum transfer from unit
mass of the stream to unit mass of injected gas the less this
contrast. At sufficiently high incident Mach number and low enough
temperature contrast, the tail material achieves supersonic velocities
in terms of the sound speed in the stream and obviously as well that
of the injected gas (since the injected gas is either at the same
temperature or cooler than the stream gas).

To compare the tail models with the observational data we assume that
a stream of velocity $V_{\rm S}$ interacts with a clump which has a velocity
$V_{\rm C}$, where both velocities are measured in the stellar frame. The
relative velocity is therefore ($V_{\rm S}-V_{\rm C}$). We concentrate
on knot~38 of \citet{Meaburn:1998}. The gas at the end of this tail
has a velocity $V_{\rm T} = 22\;\kmps$ relative to the head
\citep{Meaburn:2003}. Since this knot has a tail length-to-head
diameter ratio of around 100:1 we take this to be equal to the
coasting velocity from our calculations reached at 100 injection
radii.

We can write $V_{\rm T} = \beta(M)(V_{\rm S} - V_{\rm C}) = \beta(M)M
c_{\rm S}$ where $\beta(M)$ is a constant dependent on the Mach number
$M$ of the interaction and the stream-source sound speed ratio and
$M\equiv(V_{\rm S} - V_{\rm C})/c_{\rm S}$.  In Table~\ref{tab:beta}
we give values of $\beta(M)$ for the cases considered here.

\begin{table}
\caption[]{The value of $\beta(M)$ as a function of $M$ and 
$c_{\rm S}/c_{\rm i}$.}
\label{tab:beta}
\centering
\begin{tabular}{ccc}
\hline
\hline
M & $c_{\rm S}/c_{\rm i} = 1$ & $c_{\rm S}/c_{\rm i}=2$ \\
\hline
1 & 0.91 & 0.76 \\
2 & 0.81 & 0.54 \\
4 & 0.66 & 0.43 \\
\hline
\end{tabular}
\end{table}

In Table~\ref{tab:vz} we give tail velocities calculated for various
Mach numbers for the two temperature ratios above using the global
knot velocity of $14\;\kmps$. We use a sound speed of $11.7\;\kmps$
(i.e. $c_{\rm i} = 10^{4}\;{\rm K}$) for the source gas. We also list
values of $V_{\rm S}$, the stream velocity in the stellar frame, for
comparison with the value of $31\;\kmps$ of the overrunning stream
found by \citet{Meaburn:1998}.

The observed acceleration towards positive radial velocities along the
tail from knot~38 is shown particularly well in the line profiles in
Fig.~13 of \citet{Meaburn:1998} and in the velocity images in Figs.~4
and~12 of that paper (though first correct the error in the legend for
Fig.~12 which includes four panels whose indicated positions should
all be reversed i.e. bottom-left, bottom-right, top-left, top-right
should read top-right, top-left, bottom-right, bottom-left
respectively). With this correction it can be seen in Fig.~12 that the
tail furthest from the knot becomes more prominent in the more
positive radial velocity range.

\begin{table}
\caption[]{Tail ($V_{\rm T}$) and stream ($V_{\rm S}$) velocities ($\kmps$).}
\label{tab:vz}
\centering
\begin{tabular}{ccccc}
\hline
\hline
  & \multicolumn{2}{c}{$c_{\rm S}/c_{\rm i}=1$} & \multicolumn{2}{c}{$c_{\rm S}/c_{\rm i}=2$} \\
\hline
M & $V_{\rm T}$ & $V_{\rm S}$ & $V_{\rm T}$ & $V_{\rm S}$ \\
\hline
1 & 11 & 26 & 18 & 38 \\
2 & 19 & 37 & 25 & 61 \\
4 & 31 & 61 & 40 & 107 \\
\hline
\end{tabular}
\end{table}

We see that the tail and overrunning stream velocities are reasonably
well reproduced with $M \approx 2$, $c_{\rm S}/c_{\rm i} = 1$ and 
$M \approx 1$, $c_{\rm S}/c_{\rm i} = 2$.

Tails should show more complex velocity structure closer to the
injection zone particularly with anisotropic mass injection but we do
not consider this further here.

The setting up time for the tail of knot~38 $\approx 2-4 \times
10^{3}\;{\rm yrs}$ with a length of $2.5 \times 10^{17}\;\cm$ and a
tail gas velocity of $10-20 \;\kmps$. This implies formation after
fast wind activity has ceased. Although \citet{Patriarchi:1991} showed
from IUE observations that 60\% of central stars of PNe emit fast
particle winds, \citet{Cerruti-Sola:1985} failed to detect one from
the central star of NGC~7293. Shorter tails take proportionally
shorter times to form and more ``typical'' tails with lengths of a few
times $10^{15}\;\cm$ could be set up in timescales of the order of a
hundred years. It is not clear why the tails are generally fairly
short. One possibility might be time fluctuations in stream or source
injection properties.

\section{Clump motions}
The effects on tail production mechanisms of systematic clump motions,
radial or non-radial, have received little attention. The non-radial
component of the Helix knot expansion velocity is unclear. On general
grounds, non-radial velocity components of clumps seem inevitable if
clumps form in dynamical instabilities. If they are generated by shell
fragmentation, they should have some dispersion around the shell
velocity at fragmentation since it is highly unlikely that shells are
strictly spherical.

\citet{Diamond:2003} measured global proper motions of the SiO maser
spots in the Mira variable TX~Cam that correspond to velocities of
about $10\;\kmps$ and noted that ``significant'' non-radial velocity
components are present. These motions are in a stellar atmosphere of
radius $\sim {\rm few} \times 10^{13}\;\cm$. If these spots move out to
radial distances $\sim 10^{17}\;\cm$, conservation of angular momentum
ensures very small non-radial velocity components ($\ll 1 \;\kmps$).

In Paper~IV results are given for situations where clumps move at an
angle to a global flow. To a good approximation, well-developed tails
lie along the resultant velocity vector of the stream-clump
velocities. If we assume that the stream is strictly radial, then
tails are aligned to an angle $\alpha \approx {\rm
arc\;\;tan}[V_{\rm tan}/(V_{\rm S}-V_{\rm C})]$ with the radial direction,
where $V_{\rm tan}$ is the non-radial velocity component of the clump
velocity.  In Table~\ref{tab:vtan} we give maximum values of $V_{\rm
tan}$ that allow an alignment angle of 5$^{\circ}$ to the radial
direction where we take this angle as characteristic.

\begin{table}
\caption[]{$V_{\rm tan}$ ($\kmps$) for an alignment angle of 5$^{\circ}$.}
\label{tab:vtan}
\centering
\begin{tabular}{ccc}
\hline
\hline
M & $c_{\rm S}/c_{\rm i} = 1$ & $c_{\rm S}/c_{\rm i}=2$ \\
\hline
1 & 1 & 2 \\
2 & 2 & 4 \\
3 & 4 & 8 \\
\hline
\end{tabular}
\end{table}

The best-fit cases for $V_{\rm T}$ and $V_{\rm S}$ detailed above
require very small values of $V_{\rm tan}$ ($\ltsimm 2\;\kmps$) for
alignment.  This is unlikely to be satisfied with shell instability
formation but is more consistent with the much lower non-radial
velocity components expected if the clumps originated in stellar
atmospheres.

\section{Tail persistence}
\begin{figure*}
\centering
\begin{tabular}{l}
\psfig{figure=m2.0_c1.0_iso_dump6_rho_new.ps,width=2.5cm}
\psfig{figure=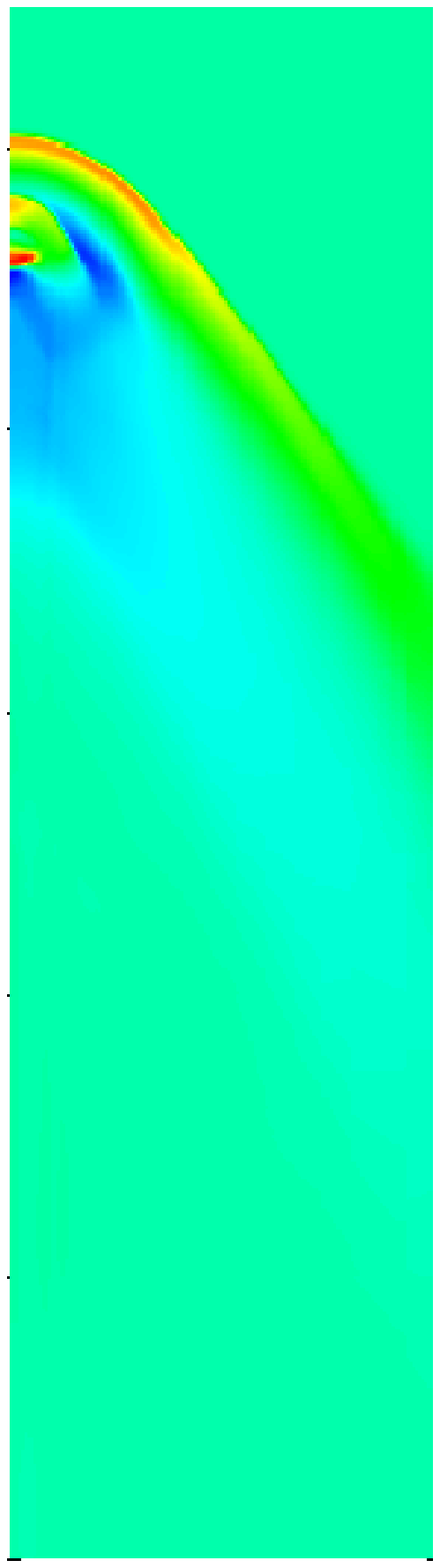,width=2.5cm}
\psfig{figure=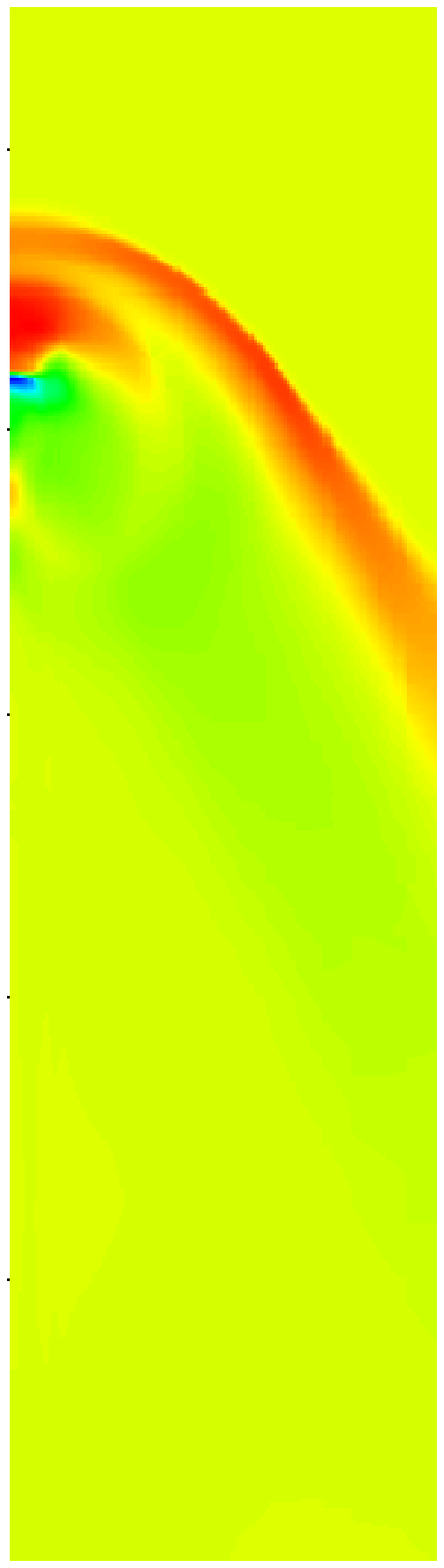,width=2.5cm}
\psfig{figure=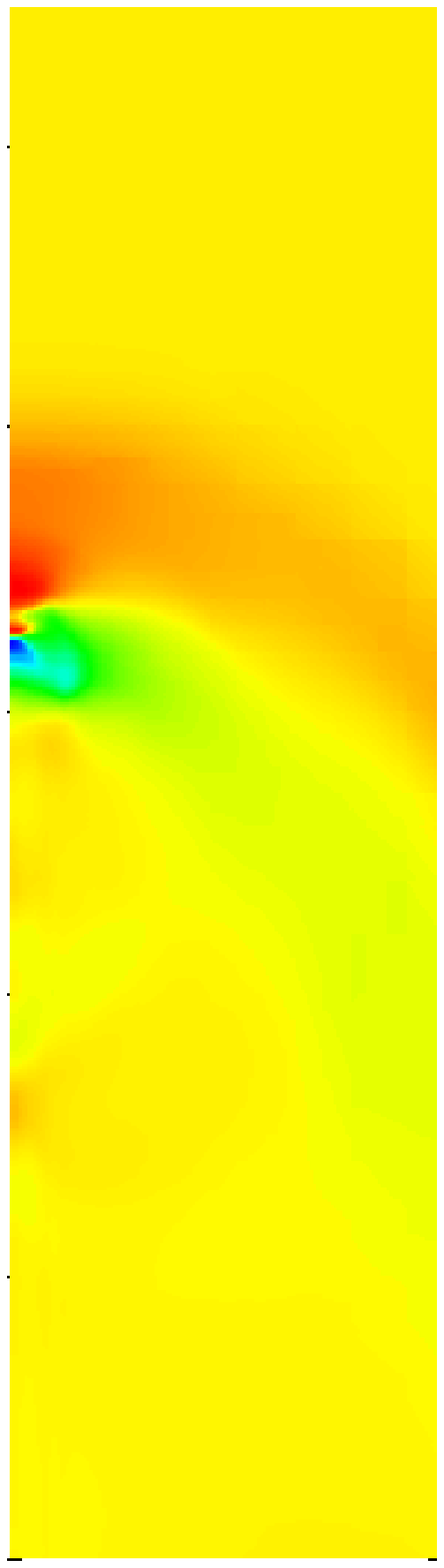,width=2.5cm}
\psfig{figure=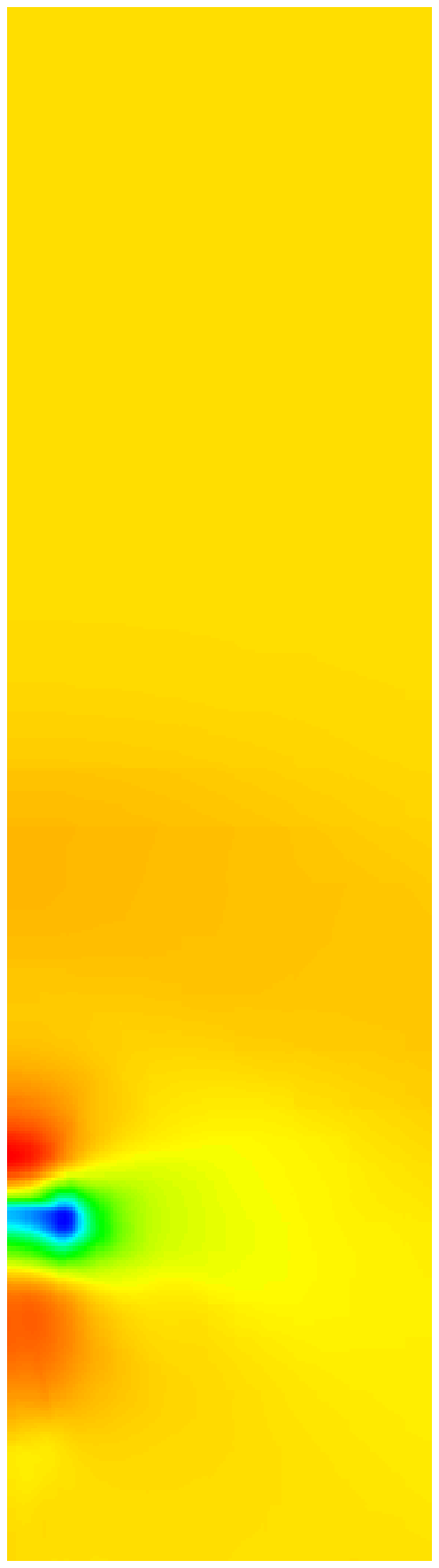,width=2.5cm}\\
\psfig{figure=m2.0_c1.0_iso_dump6_col_new.ps,width=2.5cm}
\psfig{figure=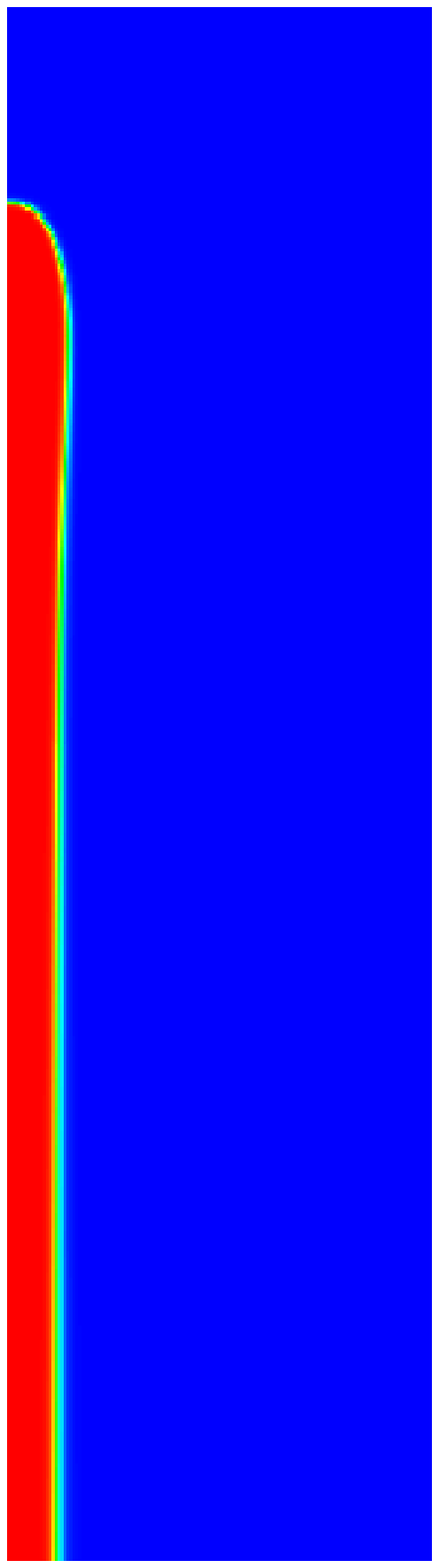,width=2.5cm}
\psfig{figure=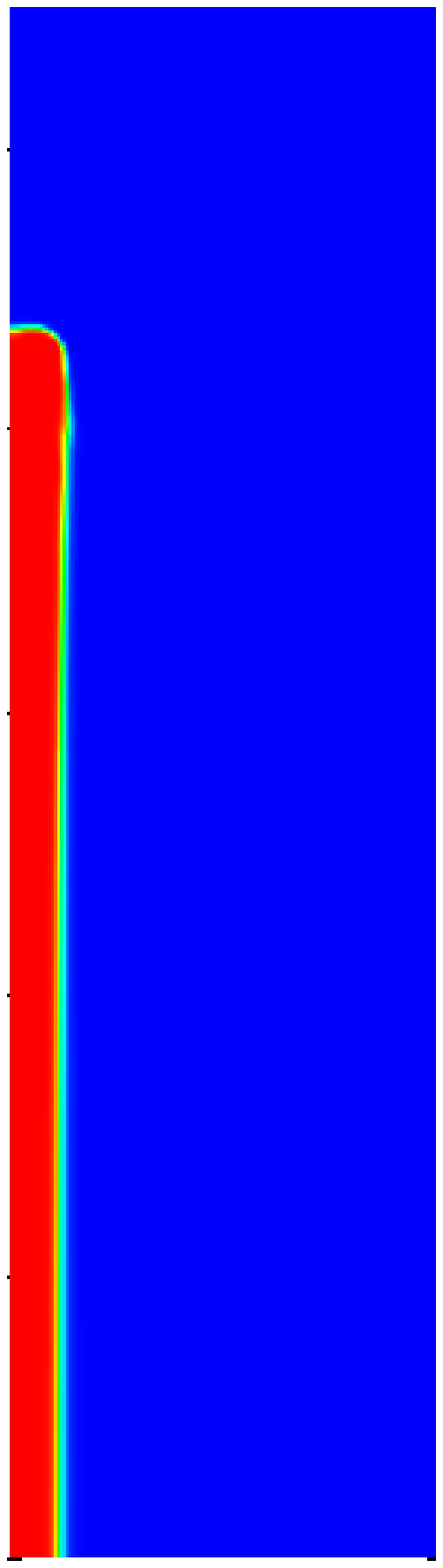,width=2.5cm}
\psfig{figure=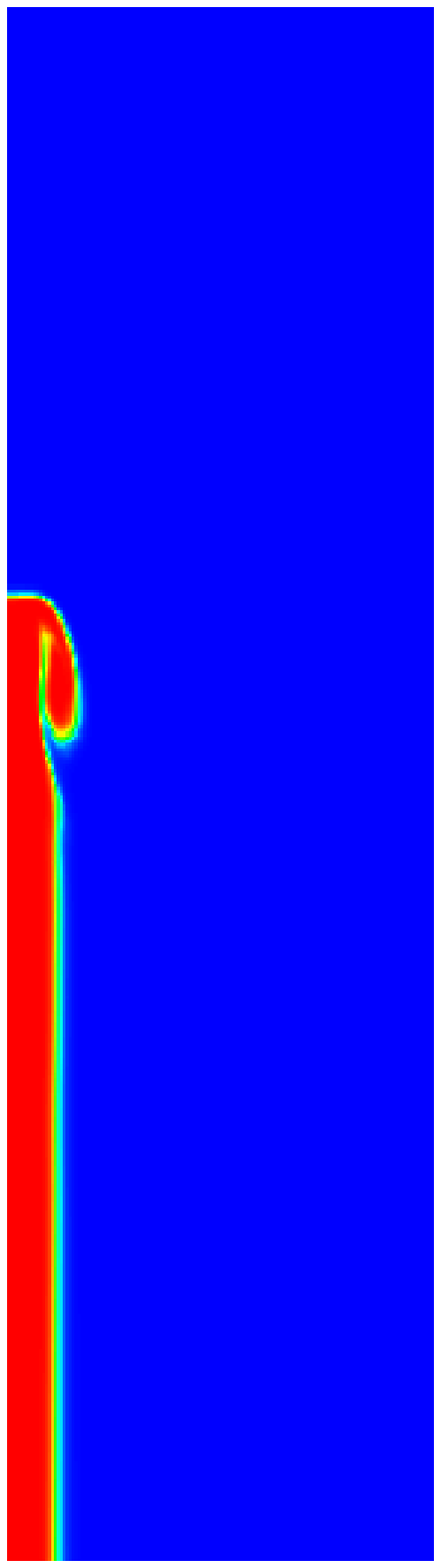,width=2.5cm}
\psfig{figure=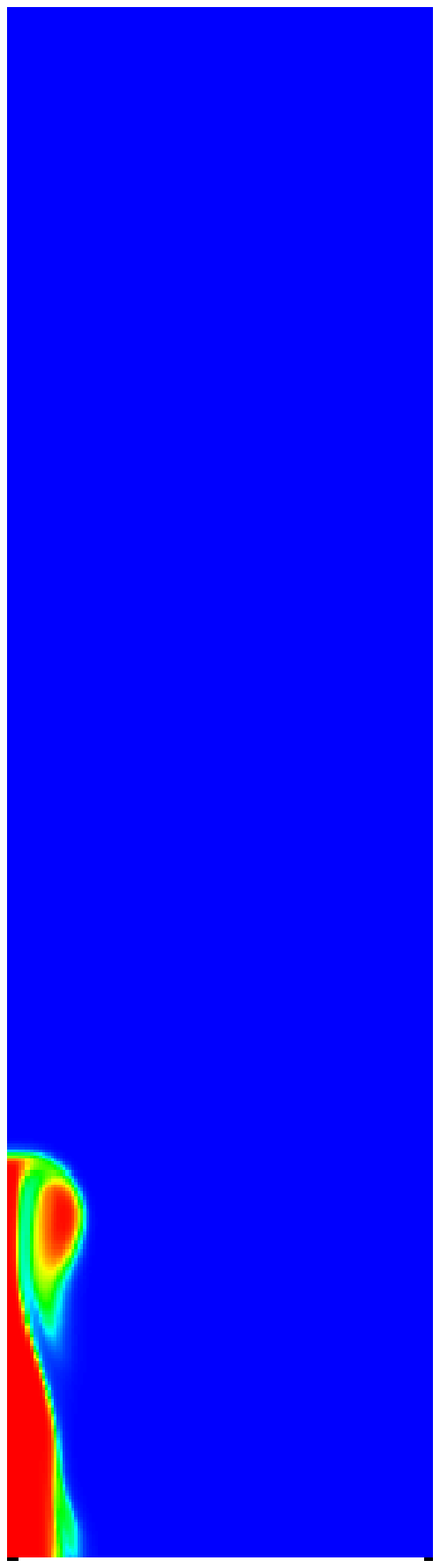,width=2.5cm}
\end{tabular}
\caption{Evolution of the $M=2$, $c_{i}/c_{s}=1.0$
isothermal-isothermal simulation after the source injection is
switched off. t=0 (left), 6.0, 13.4, 28.3, and 58.0 (right). The
isothermal sound speed=1.0 for both the injected and ambient flow, and
the ambient flow speed=2.0. The top panels show density plots, while
the bottom panels show the tracer for injected material. The colour
scaling of each density panel is different in order to best display
the structure. From left to right, ${\rm log_{10}}\; \rho = -0.97$ to
0.99, $-0.14$ to 0.28, $-0.38$ to 0.15, $-0.38$ to 0.12, and $-0.28$ to
0.08. The density enhancement relative to the stream density 
(${\rm log_{10}}\; \rho = 0.0$) declines as time evolves.}
\label{fig:evolve}
\end{figure*}

We have examined the persistence of tails where the injection is
switched off to quantify the proposal by \citet{Meaburn:2005} that
this is responsible for the production of headless spokes.  In
Fig.~\ref{fig:evolve} we show time frames for the evolution of the
tails produced for the two best fit cases in Section~\ref{sec:tailvz}
(i.e. $M \approx 2$, $c_{\rm S}/c_{\rm i}=1$ and $M \approx 1$,
$c_{\rm S}/c_{\rm i}= 2$).  As before, for the case of a sound speed
ratio of unity we give both the advected scalar plots and the density
plots. Tails persist over timescales of at least 50 or so (in
dimensionless units). The actual time unit is equal to $2r_{\rm
C}/V_{\rm F}$, where $V_{\rm F}$ is the gas velocity in the clump
frame (i.e. $V_{\rm S}-V_{\rm C}$). If we take the sound speed in the
injected gas to be $11.7\;\kmps$ as before, the flow speed in both
cases is $23.4 \;\kmps$. If we take the injection radius as about
$10^{15}\;\cm$ corresponding to a typical clump diameter, the tails
persist for times at least $\sim 1400\;{\rm yr}$. It appears plausible
therefore that headless tails are the remains of tails behind recently
destroyed clumps.

\begin{figure*}
\centering
\psfig{figure=m1.0_c0.5_iso_dump14_rho_new.ps,width=2.5cm,clip=,bbllx=220pt,bblly=139pt,bburx=363pt,bbury=668pt}
\psfig{figure=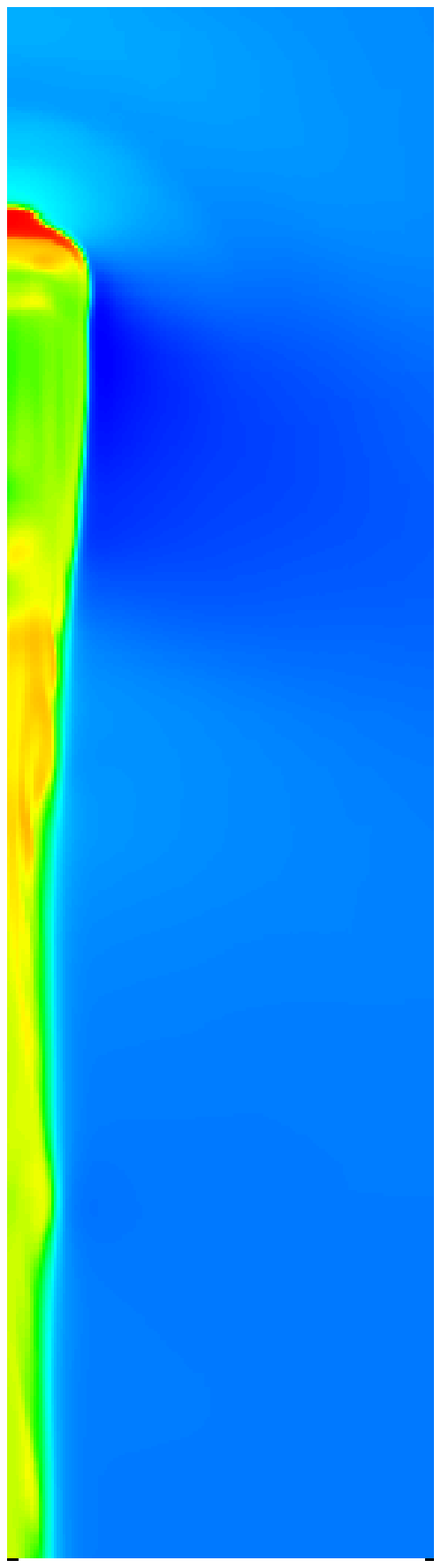,width=2.5cm,clip=,bbllx=220pt,bblly=139pt,bburx=363pt,bbury=668pt}
\psfig{figure=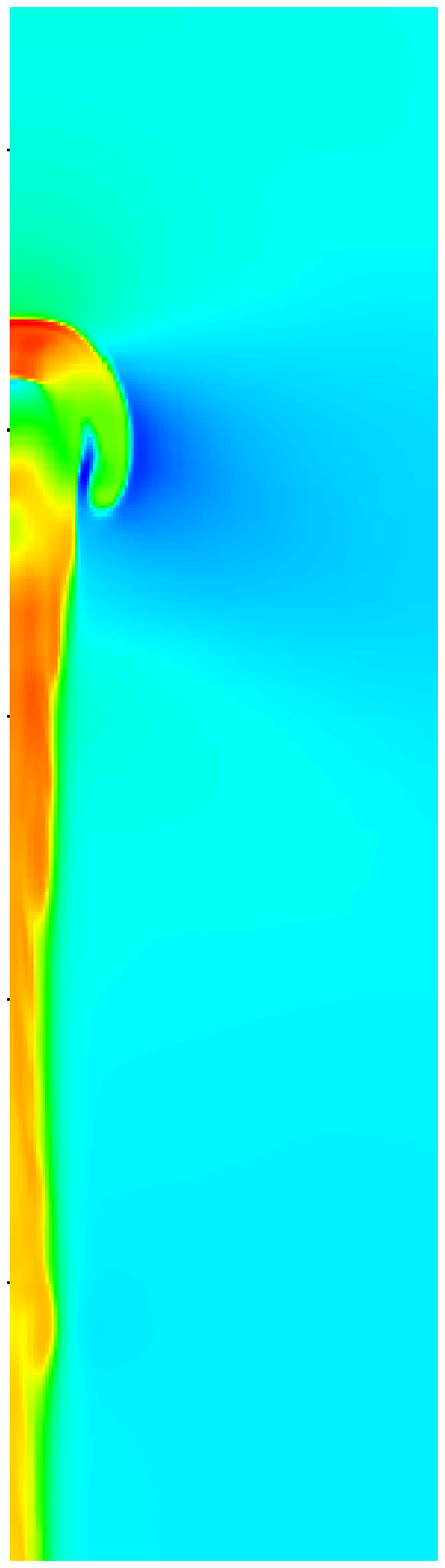,width=2.5cm,clip=,bbllx=220pt,bblly=139pt,bburx=363pt,bbury=668pt}
\psfig{figure=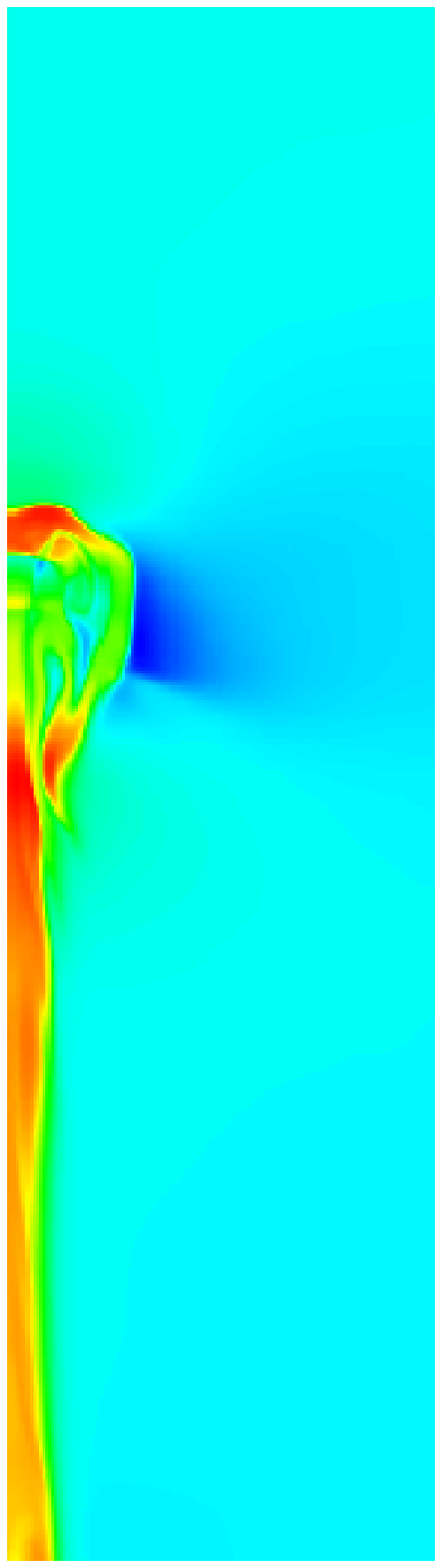,width=2.5cm,clip=,bbllx=220pt,bblly=139pt,bburx=363pt,bbury=668pt}
\psfig{figure=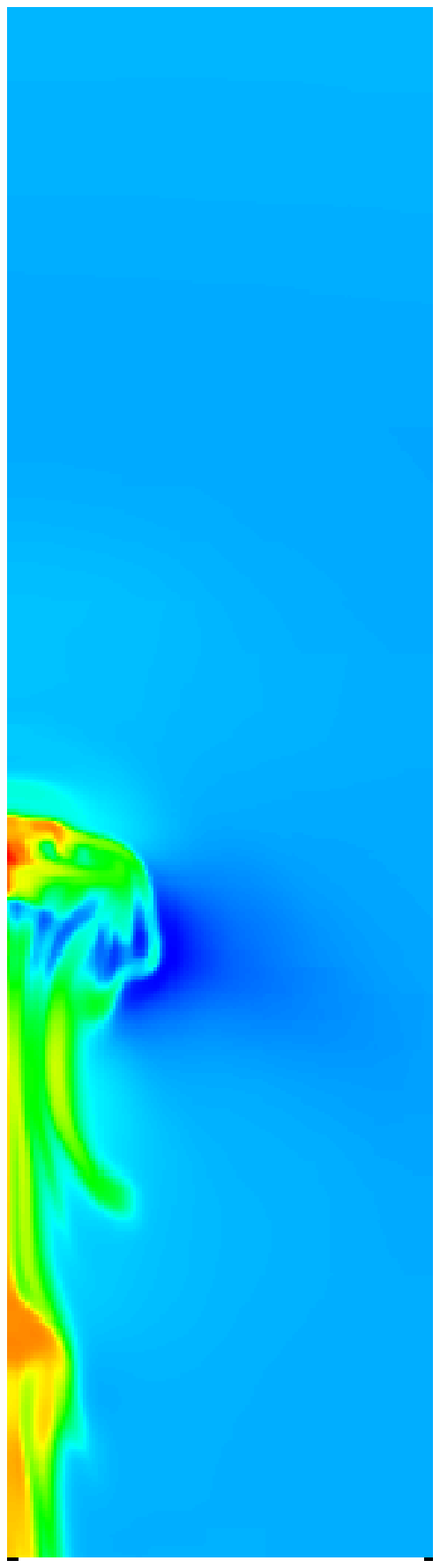,width=2.5cm,clip=,bbllx=220pt,bblly=139pt,bburx=363pt,bbury=668pt}
\psfig{figure=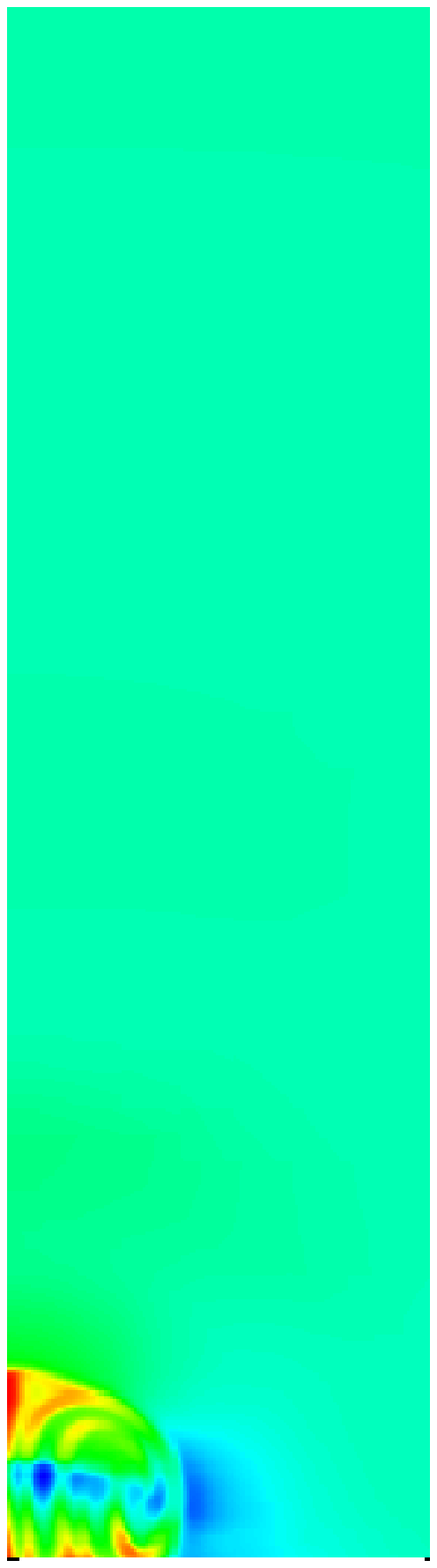,width=2.5cm,clip=,bbllx=220pt,bblly=139pt,bburx=363pt,bbury=668pt}
\caption[]{Evolution of the $M=1$, $c_{\rm S}/c_{\rm i}=2$
isothermal-isothermal simulation after the source injection is
switched off. $t=0$ (left), 10.9, 21.7, 43.4, 76.4, and 120.4
(right). The isothermal sound speed=1.0 and 2.0 for the injected and
ambient flow (respectively), and the ambient flow speed=2.0.
 The colour scaling of each panel is different in order to best
display the structure. From left to right, ${\rm log_{10}}\; \rho = -0.69$ to
0.39, $-0.70$ to 0.11, $-0.81$ to 0.04, $-0.81$ to 0.03, $-0.74$ to 0.04, 
and $-0.74$ to $-0.32$. The density enhancement relative to the stream 
density (${\rm log_{10}}\; \rho = -0.60$) declines as time evolves.}
\label{fig:evolve2}
\end{figure*}

\section{Conclusions}
Linear structures have been observed in several planetary
nebulae. This morphology has led to two classes of models. However
morphology itself is not sufficient to discriminate between
them. Velocity structure is the key and only for two knots (14 and 38)
in the Helix nebula is the necessary data available. The acceleration
along the tail associated with the spectacular knot~38 suggests
strongly that there is momentum transfer between a directed flow and a
mass source. The advanced evolutionary state of the central star, and
the details of the nebular structure, rule out any hypersonic stellar
wind-mass source interaction. Although attractive in many respects,
the interaction with a subsonic hot stream of shocked stellar wind gas
is also ruled out because the existence of a reservoir of hot gas
cannot be reconciled with the nebular structure. Much more plausible
is tail formation taking place when a transonic or moderately
supersonic stream of photoionized gas overruns gas injected by the
photoionization of cometary globules. The injected gas is accelerated
by momentum transfer from the stream and the velocities that are
generated are in good agreement between those measured for the likely
overrunning stream and the tail gas of knot~38 in the Helix. Thus the
general scheme for the formation of cometary globules involves flows
(or winds) overrunning a slowly expanding system of pre-existing dense
neutral globules. At least for NGC~7293, the knotty nature of the
ambient neutral shells associated with this object surely precludes
any other possibility (instabilities in a global expanding shell for
example). In fact the very strong radial alignment of the tails
suggests that when overrun, the globules have a velocity that can at
most have a very small non-radial component. This is consistent with
density structures that originate at radii far less that the present
scale size of the nebula. Finally, it also appears likely that the
widely distributed striations seen in the nebular gas represent the
remains of tail structures generated by mass injection from now
destroyed globules.

\begin{acknowledgements}
JMP gratefully acknowledges support from the Royal Society.  
This research has made use of NASA's Astrophysics Data System Abstract Service.
\end{acknowledgements}

\end{document}